\pdfoutput=1 
\documentclass[10pt, aps, prb, twocolumn, showpacs, superscriptaddress,longbibliography]{revtex4-1}
\usepackage{lmodern}
\usepackage{amsmath}
\usepackage{graphicx}
\usepackage{xcolor}
\usepackage{natbib}
\usepackage{physics}
\usepackage[colorlinks,bookmarks=false,citecolor=blue,linkcolor=blue,urlcolor=blue,hyperfootnotes=true]{hyperref}

\begin{document}
\title{Toward flying qubit spectroscopy}

\author{Beno\^it Rossignol}
\author{Thomas Kloss}
\author{Pac\^ome Armagnat}
\author{Xavier Waintal}
\affiliation{Univ.\ Grenoble Alpes, CEA, INAC-Pheliqs, 38000 Grenoble, France}

\date{February 16, 2018}

\begin{abstract}
While the coherent control of two level quantum systems ---qubits--- is now standard, their continuum electronic equivalents ---flying qubits--- are much less developed. A first step in this direction has been achieved in DC interferometry experiments. Here, we propose a simple
setup to perform the second step, the spectroscopy of these flying qubits, by measuring the DC response to a high frequency AC voltage
drive. Using two different concurring approaches --- Floquet theory and time-dependent simulations --- and three different models --- an analytical model, a simple microscopic model and a realistic microscopic model ---  we predict the power-frequency map of the multi-terminal device. We argue that this spectroscopy provides a direct measurement of the flying qubit characteristic frequencies and a key validation for more advanced quantum manipulations.
\end{abstract}
\maketitle

\section{Introduction}
The development of a new type of quantum bit happens in stages.
Let us consider the singlet-triplet double quantum dot qubit \cite{Loss1998} as a typical example. In this case, the first stage consists of DC measurements of the so-called stability diagram. 
Once a suitable physical regime has been found, stage II consists of performing the spectroscopy of the qubit to assert its suitability  and determine its dynamical characteristics. This can be done through, e.g. electronic dipolar spin resonance (EDSR) \cite{Zavoisky}.
It is crucial to pass these two stages before one can consider sending more elaborated pulse sequences like Rabi, Ramsey and echo experiments. In the last stage (before considering coupling several of these qubits), one implements single shot measurements. 

Quantum mechanics, however, is not limited to bound states and propagating quantum states instead of bound states could also be used to form qubits. The so-called flying qubits have been successfully realized with photons in linear quantum optics \cite{Knill2001,Maxein2013} but here we focus on proposals based on electrons \cite{BAUERLE2018}. The first stage of the electronic flying qubit \cite{Haack2011,Fve2008} implementation has been demonstrated in several experiments that show controlled two paths interferometry  in two dimensional electron gas in the presence \cite{Ji2003,Roulleau2008} or absence \cite{Bautze2014,Yamamoto2012} of magnetic field as well 
as in graphene \cite{Wei2017}.
Other features, specific to propagating quantum systems, have also been demonstrated (including single electron sources \cite{Dubois2013,Bocquillon2013,McNeil2011,Feve2007} and their Hong-ou-Mandel characterization) or proposed theoretically \cite{Gaury2014b,Hofer2014,Gaury2015}. However stage II, the spectroscopy of a flying qubit, has not yet been realized experimentally.

The electronic flying qubits that we consider in this article are "two paths" interferometers, the electronic analogue of the Mach-Zehnder interferometer sudied in optics. The two states of the qubits are coded in the two paths $\uparrow$ or $\downarrow$ that a single electronic excitations use for propagation. Here the role of the qubit frequency is replaced by $\hbar/\tau$ where $\tau$ is a characteristic time,
a difference between two times of flight (to be defined below), of the device. Similarly to localized system that may have multiple energy levels, there may be several propagating
channels giving rise to several characteristic times $\tau$. Measuring these times and
assessing that electronic interferometry experiments can be performed at high frequency is the next key milestone of the field.
 
In this article, we propose to use quantum rectification (measurement of a DC current in presence of a high frequency sinusoidal drive)\cite{Polianski2001, Brouwer2001, Vavilov2005,Arrachea2006,Giblin2013,Vavilov2005,MartnezMares2004}  as a tool to perform the spectroscopy of flying qubits. We argue that quantum rectification provides a clear spectroscopy of the device while being much more accessible experimentally than other techniques, in particular in the challenging $\sim$ 10\,GHz\,$-$\,1\,THz frequency range which is required for this type of physics.

\section{A two paths electronic interferometer using a split wire geometry}
\begin{figure}[ht]
\centering
\includegraphics[scale=0.3]{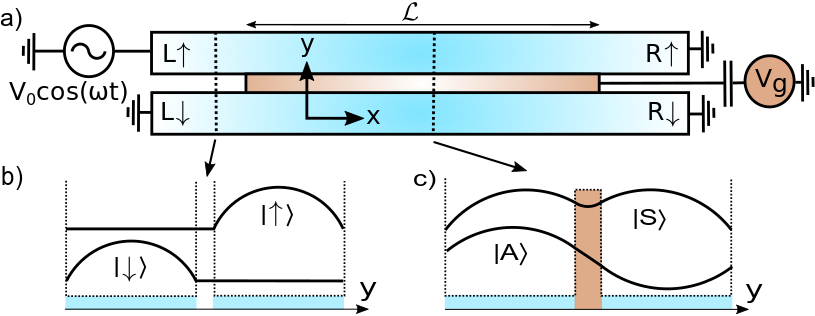}
\vspace{-5ex}
\caption{(Color online) Upper panel: schematic of the flying qubit geometry. Two wires labeled $\uparrow$, $\downarrow$ are connected to two electrodes on the left (L$\uparrow$, L$\downarrow$) and two electrodes on the right (R$\uparrow$, R$\downarrow$). Lower panel: schematic of the transverse part of the propagating modes close to the electrodes (left) and in the central tunneling region (right).}
\label{fig1}
\end{figure}

We focus this study on the tunneling wire  ``flying qubit'' geometry sketched in Fig.\,\ref{fig1}a and studied experimentally in \cite{Yamamoto2012,Bautze2014,Takada2014,Takada2015}. 
The device consists of two quasi-one dimensional wires labeled 
$\uparrow$ (upper) and $\downarrow$ (lower) connected to four electrodes: two on the left L$\uparrow$, L$\downarrow$ and two on the right 
R$\uparrow$, R$\downarrow$. Close to the electrodes, the wires are disconnected. However, in a central region of length $\mathcal{L}$,
the two wires are in contact so that an electron can tunnel back and forth from the upper to the lower part. A capacitive top gate $V_g$ 
controls the intensity of the tunneling coupling between the wires. The coherent oscillation that takes place in the tunneling region between the upper and lower wire can be interpreted as a quantum gate operated on the flying qubit. Equivalently, an electron entering the upper wire decomposes into a superposition of a symmetric and antisymmetric propagating states which forms a two-path interferometer.

The DC characteristics of this device have been analyzed previously \cite{Bautze2014,Weston2016a} both theoretically and experimentally. For completeness, we recall here its salient features. Let us determine the scattering matrix of this device in the limit where (i) there is only one propagating channel in each of the wires and (ii) the spatial variation of the tunneling coupling is very smooth with respect to the Fermi wave length. This implies that there is no reflection in the device as backscattering involves the $2k_F$ Fourier component of the
potential ($k_F$ is the Fermi momentum): an electron injected on the left, say in L$\uparrow$ is transmitted either toward R$\uparrow$ or R$\downarrow$. To determine the transmission amplitude $d_{ba}(E)$ from channel $a$ on the left to channel $b$ on the right ($a,b \in\{ \uparrow,\downarrow\}$), let us consider
the transverse part of the propagating modes. A schematic representation of these wavefunctions is shown in Fig.\,\ref{fig1}b for the decoupled wires (close to the electrodes) and in Fig.\,\ref{fig1}c for the tunneling region. In the latter, the $\uparrow$ and $\downarrow$ channels 
hybridize into a symmetric $S$ and an antisymmetric $A$ channels of respective longitudinal momentum $k_S$ and $k_A$ along the $x$ direction.
The key point is to recognize that the $S\ (A)$ channel is continuously connected to the symmetric (anti-symmetric) combination of the $\uparrow$ and $\downarrow$ channels,
$\ket{S/A} \leftrightarrow \left( \ket{\uparrow} \pm \ket{\downarrow} \right) / {\sqrt{2}}.
$
Hence an electron injected in $\ket{\uparrow}$,
\begin{equation}
\ket{\uparrow} = \frac{1}{2}\left( \ket{\uparrow} + \ket{\downarrow} \right) +\frac{1}{2} \left( \ket{\uparrow} - \ket{\downarrow} \right) \rightarrow \frac{1}{\sqrt{2}} (\ket{S} +\ket{A}) ,
\end{equation} 
is transmitted into S and A with amplitude $1/\sqrt{2}$.
Inside the tunneling wire, the wavefunction picks up a phase $e^{i\phi_{S/A}}$ which in the WKB approximation reads $\phi_{S/A} = \int_0^\mathcal{L} dx \ k_{S/A}(x) \approx k_{S/A}\mathcal{L}$. After the tunneling region, the $S$ and $A$ recombine into the $\uparrow$ and $\downarrow$ channels and we arrive at,
\begin{equation}
d_{\uparrow\uparrow}(E) =\frac{1}{2}(e^{i\phi_S}+e^{i\phi_A}), \,\,
d_{\downarrow\uparrow}(E) =\frac{1}{2}(e^{i\phi_S}-e^{i\phi_A}).
\label{DCd}
\end{equation}
The differential conductance $g_{ba}$ that relates the current flowing on the right in lead $b$ from an increase of voltage in the left
on lead $a$ is given by the Landauer formula, $g_{ba} = (e^2/h) D_{ba}(E_F)$ with $ D_{ba}(E_F) = |d_{ba}(E_F)|^2$ and $E_F$ the Fermi energy (we ignore spin everywhere; it can be restored by simply multiplying the currents by a factor 2).
The above analytic expressions have been shown to grasp the important features of the corresponding experimental devices in DC \cite{Bautze2014}. In particular, upon decreasing the gate voltage $V_g$ toward large negative values, $k_S-k_A$ decreases towards zero (the two channels become increasingly alike) and the differential conductance  $g_{\uparrow\uparrow} \propto \cos^2((\phi_S-\phi_A)/2)\approx \cos^2 ((k_S-k_A)\mathcal{L}/2)$ first oscillates, then saturates to perfect transmission.

For the AC response discussed in this article, we need the energy dependence of the transmission amplitude. Linearizing the dispersion relation of the $S$ and $A$ channels, we introduce the corresponding velocity $v_{S,A} = (1/\hbar) dE_{S,A}/dk$ and the time of flight 
$\tau_{S/A}= \mathcal{L}/v_{S,A}$ through the channel. The phase difference $\phi_S(E) - \phi_A(E)$ is controlled by
the difference $\tau \equiv \tau_S-\tau_A $ of time of flight, and we arrive at 
\begin{equation}
\phi_S(E) - \phi_A(E) \approx \delta_F +  (E-E_F) \tau/\hbar
\label{linearphi}
\end{equation} 
with $\delta_F \equiv \phi_S(E_F) - \phi_A(E_F)$.

\section{A General formula for calculating rectification currents}
We now develop the scattering theory of the rectified direct current generated by an AC voltage drive.
We consider a multiterminal mesoscopic system and apply a periodic time dependent voltage $V(t)$ to one electrode 
(for definiteness, we focus below on L$\uparrow$) with frequency $\omega$.
We seek to obtain the average (over time) DC current flowing in the different electrodes. Such a calculation can be performed in different
but fully equivalent ``Floquet'' formalisms including the scattering \cite{Moskalets2011}, Non-equilibrium Green's function \cite{Shevtsov2013} or wave function approach \cite{Gaury2014}. Here, we follow the latter after Ref.\,\cite{Gaury2014, Gaury2015}.

In what follows, we neglect the spatial dependence of the electric potential drop, i.e. we suppose that the drop of electric potential takes place very abruptly at the Ohmic contact - two dimensional gas interface. Such an approximation is well justified in the present case due to the presence of the electrostatic gates that define the conducting region. These gates are metallic, hence equipotential; they ensure that the potential drop takes place on a distance which is essentially set by the distance between the gate and the two-dimensional electron gas. This distance is typically of the order of 100 nm which is much shorter than the size of the device (typically 10 $\mu m$) so that the approximation of perfectly sharp drop is reasonably accurate. In the opposite situation (absence of electrostatic gates) the potential drop would be linear between the two contacts. A discussion of this problem can be found in section 8.4 of \cite{Gaury2014}.
The abrupt drop of potential is an important ingredient for the physics of propagating pulses such as the minimum excitations "Levitons". The recent experiments that measured the time of flight of such pulses \cite{Roussely2018} provide a clear experimental evidence that the drop is indeed sharp and take place at the Ohmic contact - electronic gas interface, since well defined velocities could be measured.

The effect of the time dependent voltage is to dress an incoming wave function of the form $e^{ikx -iEt/\hbar}$ with an extra phase factor 
$e^{-i\Phi(t)}$ [with $\Phi (t) \equiv \int_0^t dt'\  eV(t')/\hbar$] that accounts for the variation of electric potential.
Decomposing this phase into its Fourier component $P_n$,
\begin{equation}
e^{-i\Phi(t)} = \sum_{n} P_n e^{-i\omega n t},
\end{equation}
the net effect of $V(t)$ is that the incoming wave function is now a coherent superposition 
$\sum_{n} P_n e^{ikx-iEt/\hbar-i\omega n t}$ of plane waves at different energy. As different energies get transmitted
into different channels, we arrive at the following time dependent transmission amplitude for an incoming energy $E$,
\begin{equation}
d_{ba}(t,E) = \sum_{n} P_n d_{ba}(E+n\hbar\omega) e^{-iEt/\hbar-i\omega n t}
\end{equation} 
where $d_{ba}(t,E)$ is the Fourier transform with respect to $E'$ of $d_{ba}(E',E)$ which is itself the inelastic amplitude to be transmitted from energy $E$, lead $a$ toward energy $E'$, lead $b$.
The generalization of the Landauer formula to time dependent currents provides the time dependent current $I_b(t)$ as
\begin{equation}
I_b(t) =\frac{e}{\hbar} \int \frac{dE}{2\pi} \left[|d_{ba}(t,E)|^2 - |d_{ba}(E)|^2\right] f_a(E) 
\end{equation} 
where $f_a(E)$ is the Fermi function of the lead $a$ subject to the time dependent voltage. The second term in the previous equation subtracts the current sent from lead $a$ in the absence of time-dependent  voltage which is a convenient way to ensure the overall current conservation\cite{Gaury2014,Gaury2014b}. 
Focusing on the DC (rectification) current $\bar I_b = \omega/(2\pi) \int_0^{2\pi/\omega} dt I_b(t)$ we arrive at,
\begin{equation}
\label{semi-analytical}
\bar I_b = \frac{e}{h} \sum_n |P_n|^2 \int dE|d_{ba}(E)|^2\left[f_a(E+n\hbar\omega)-f_a(E)\right].
\end{equation} 
Equation (\ref{semi-analytical}) is very general and relates the rectification properties of an arbitrary mesoscopic system to
its scattering matrix $d_{ba}(E)$, a well known DC object. In particular, it can be easily evaluated numerically for a large class of microscopic models using readily available numerical packages (in our case the Kwant \cite{Groth2014} package) for
arbitrary periodic pulses. 
We note that following the same arguments as Ref.\,\cite{Gaury2014}, we find that the rectification current is ``conserved'' and ``gauge invariant'' in the sens defined by B\"uttiker \cite{Bttiker1993}, i.e. the DC current in electrode $a$ is exactly compensated by the DC currents in the other leads and applying an AC potential on all the leads simultaneously does not generate any DC current.

\begin{figure}[t]
\centering
\includegraphics[scale=0.60]{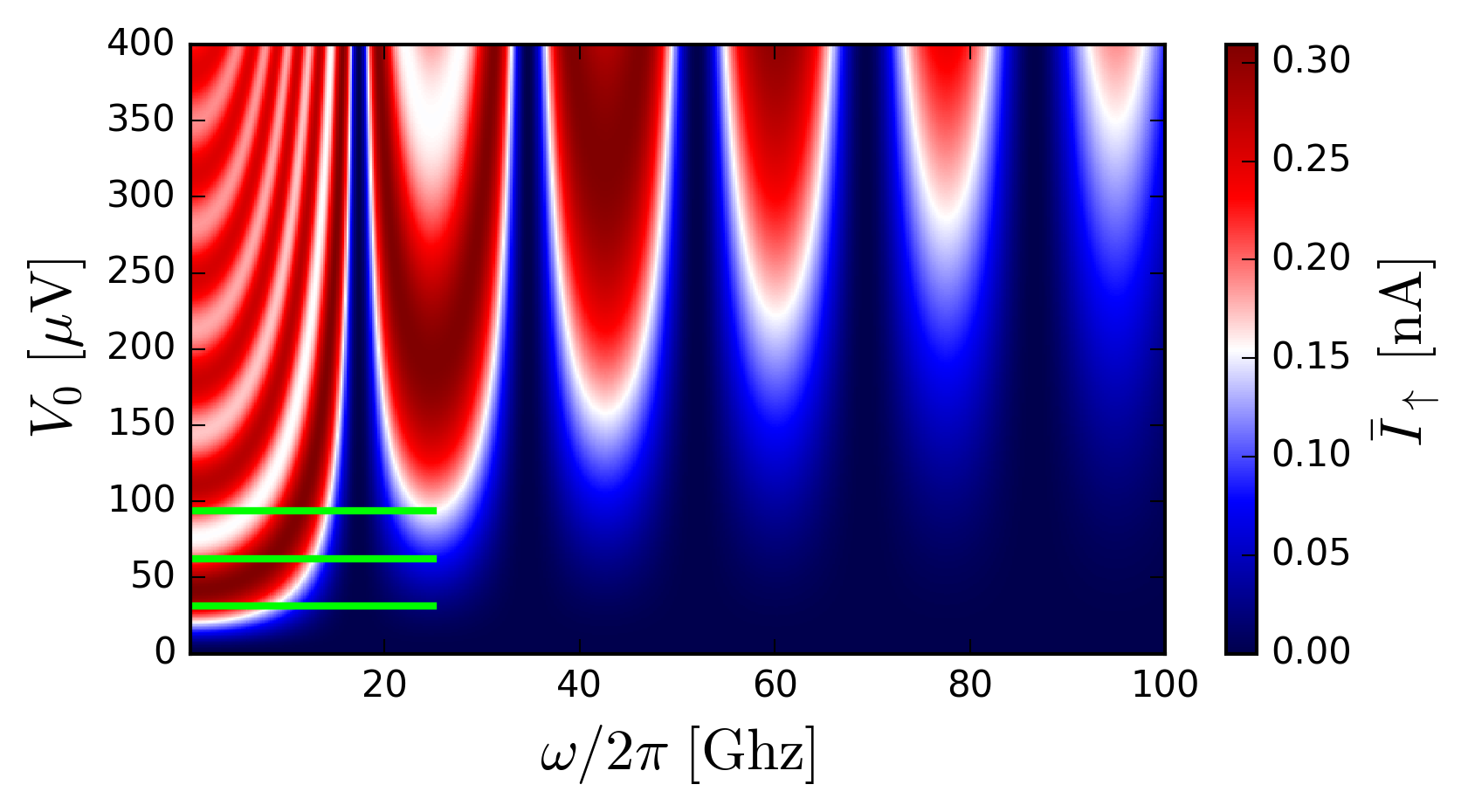}
\vspace{-3ex}
\caption{(Color online) Rectified DC current from Eq.\,(\ref{Ianalytic}) for $\delta_F=0.32\pi$ and $\tau=58$\,ps. The results of Fig.\,\ref{fig4} correspond to cuts along the green lines. 
\label{fig2}}
\end{figure}

\section{Application to the flying qubit}

\subsection{Simple scattering model}
We now make a specific calculation using our analytical model Eq.\,(\ref{DCd}) for the flying qubit geometry. We also specialize to a drive $V(t) = V_0 \cos\omega t$ with a unique frequency which implies $P_n = J_n(eV_0/\hbar\omega)$ where $J_n(x)$ is the Bessel function of the first kind. Up to an irrelevant phase factor, the time dependent transmission reads,
\begin{equation}
d_{\uparrow\uparrow}(t,E) = \frac{1}{2} \left[ 1 + e^{i\delta_F +i\tau (E-E_F)/\hbar} e^{i\Phi(t)} e^{-i\Phi(t-\tau)}\right].
\end{equation} 
Following the same route as in the general case, and assuming zero temperature for simplicity, we get,
\begin{subequations}
\begin{align}
\label{Ianalytic}
\bar I_{\uparrow}&=\frac{e}{4\pi\tau}\sin(\delta_F)\left[J_0\left(\frac{2eV_0}{\hbar\omega}\sin(\frac{\omega\tau}{2})\right)-1\right]\\
\bar I_{\downarrow}&=-\bar I_{\uparrow}.\label{Ianalytic2}
\end{align}
\end{subequations}
Eqs.\ (\ref{Ianalytic}, \ref{Ianalytic2}) call for a few comments. (i) Even though we apply the oscillatory voltage on the upper left electrode, no DC current actually flows there as implied by Eq.\ (\ref{Ianalytic2}) and current conservation. Instead, the DC rectified current is pumped from the upper right to the lower right electrode. (ii) Eq.\,(\ref{Ianalytic}) is non perturbative both with respect to frequency and drive amplitude.
An illustrative color plot is shown in Fig.\,\ref{fig2}. It shows rich oscillatory features both as a function of $\omega$ and $V_0$. Fig.\,\ref{fig2} is the flying qubit analogue of the usual spectroscopy maps. (iii) The adiabatic limit $\omega \rightarrow 0$ can be understood without using the time dependent Floquet formalism. First, we compute the DC current-voltage characteristics 
\begin{align}
I(V) =& (e/h)\int_{E_F}^{E_F+eV} dE\  |d_{\uparrow\uparrow}(E)|^2\nonumber\\
=&\frac{e^2}{2h} V + \frac{e}{2\pi\tau} \sin (\frac{eV\tau}{2\hbar}) \cos (\delta_F + \frac{eV\tau}{2\hbar}).
\end{align}
Then the adiabatic rectified DC current is found by computing the time average of $I(V=V_0 \cos \omega t)$ and we arrive at
\begin{equation}
\label{Iadiabatic}
\bar I_{\uparrow}=\frac{e}{4\pi\tau} \sin(\delta_F)\left[J_0\left(\frac{eV_0\tau}{\hbar}\right)-1\right]
\end{equation}
which corresponds to the $\omega \rightarrow 0$ limit of Eq.\,(\ref{Ianalytic}). The rectified current is directly linked to the presence of the non-linear term in the $I(V)$ characteristics. (iv) At large $x$, the Bessel function decreases as $J_0(x) \sim \sin (x+\pi/4) \sqrt{2/\pi x}$ so that  the rectified current reaches its maximum value $\bar I_{\uparrow}=-\frac{e}{4\pi\tau_F} \sin(\delta_F)$ at large voltage and $\omega\tau = \pi$.

\subsection{Simple microscopic model} 
We now introduce a microscopic model for the Mach-Zehnder interferometer of 
Fig.\,\ref{fig1} and discuss our direct method
to perform time dependent simulations of the device. We shall find a perfect match between our time dependent simulations and a semi-analytical approach that uses the microscopic model to calculate the DC scattering matrix (using the Kwant package \cite{Groth2014}) and Eq.\,(\ref{semi-analytical}) to relate the latter to the rectified current in presence of an AC drive. 
We model the Mach-Zehnder interferometer through the following Hamiltonian,
\begin{align}
\hat{H}(t)=& \sum_{a \in \{\uparrow,\downarrow\}} \sum_{i =-\infty}^{+\infty}  [ -c_{i+1,a}^\dagger c_{i,a} + U_i c_{i,a}^\dagger c_{i,a}] \nonumber \\
&+\sum_{i =-L/2}^{+L/2} \gamma_i c_{i,\uparrow}^\dagger c_{i,\downarrow} + h.c.
\label{eq:H}
\end{align}

where $c_{i,a}$ ($c_{i,a}^\dagger$) is the usual fermionic destruction (creation) operator on site $i$ and wire $a\in \{\uparrow,\downarrow\}$. $U_i$ is an electric potential present in the central region, $\gamma_i$ characterizes the tunneling between the upper and lower wire and is controlled by the voltage $V_g$ and $L$ is the total length of the tunneling part of the wire. The nearest neighbor hopping amplitude is set to unity which defines our energy and time units ($\hbar=1$). 

For our simulations, we choose $L= 500$ sites, $E_F = 1.3$, $\gamma_i$ interpolates smoothly (over $50$ sites) between $0$ in the electrodes  
and $-0.7$ in the tunneling region. The potential $U_i$ interpolates smoothly between $0.8$ in the electrode, $1$ in a small region just before and after the tunneling region (this region is present for numerical convenience, see section 10 of Ref.\,\cite{Gaury2014}) and
vanishes inside the tunneling region. $U_i$ also includes a uniform contribution $V_0 \cos(\omega t)$ for all sites in the upper left electrode and $t>0$. For these parameters, we find a characteristic time $\tau \approx 58$ and $\delta_F \approx 0.32\pi$. These two values can be determined consistently from three different calculations: from the propagation of a voltage pulse in the time dependent simulation, from the energy dependence of the DC conductance or from the WKB approximation.

\begin{figure}[t]
\centering
\includegraphics[scale=0.45]{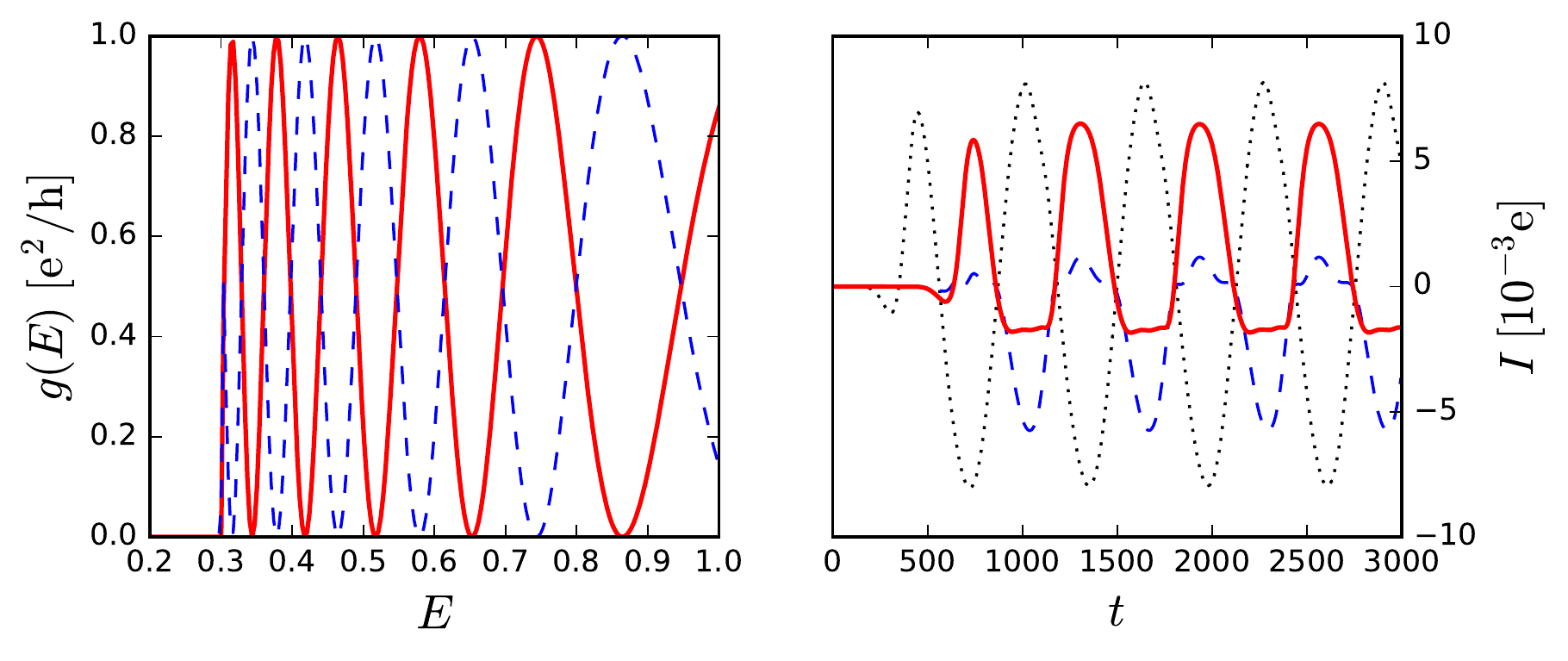}
\vspace{-3ex}
\caption{(Color online) Simple microscopic model. Left panel: DC differential conductances $g_{\uparrow\uparrow}(E)$ (dashed blue line) and 
$g_{\downarrow\uparrow}(E)$ (straight red line) as obtained from a direct numerical calculation of the tight-binding model. The numerical calculations were performed with the Kwant \cite{Groth2014} package. The reflection probability from L$\uparrow$ to L$\uparrow$ or L$\downarrow$ vanishes in the region of interest. Right panel: currents in  R$\downarrow$ (straight red line, $I_{\downarrow}(t)$) and R$\uparrow$ (dashed blue line,  $I_{\uparrow}(t)$) after a microwave excitation in L$\uparrow$ (dotted black line) computed using time dependent simulations of the microscopic model.}
\label{fig3}
\end{figure}

The time-dependent simulations are performed using the method described in Ref.\,\cite{Gaury2014,Weston2016} where all details are provided. In this method, we directly integrate the Schr\"odinger equation
\begin{equation}
i\hbar\partial_t |\Psi (t) \rangle = \hat{H}(t) |\Psi (t) \rangle
\end{equation}
without further approximations. The main difference with Eq.(\ref{semi-analytical}) lies in the treatment of the oscillatory
AC potential: in the scattering matrix approach, it is assumed that the AC potential drop does not create any back scattering.
This approximations is usually very good, up to small deviations $\sim V_0/E_F$ that have been calculated in Ref.\cite{Gaury2014}.
The left panel of Fig.\,\ref{fig3} shows an example of the DC differential conductances $g_{\uparrow\uparrow}(E)$ and  $g_{\downarrow\uparrow}(E)$ as obtained from a direct numerical calculation of the tight-binding ``simple microscopic model''.
We indeed observe the oscillations with energy discussed after Eq.\,(\ref{DCd}). We checked that the period of these oscillations matches the WKB result that can be calculated independently.
The right panel of Fig.\,\ref{fig3} shows the result (current $I(t)$ versus time $t$) of a typical time dependent simulation of the model in presence of the AC drive 
(smoothly switched on at $t=0$). These curves are averaged over time to calculate the DC rectification current $\bar I$.
 
\subsection{Comparison between the different approaches}
\label{comparison approach}
The first remarkable feature of the rectified current is the fact that it is pumped 
between the two right electrodes. The DC current in electrode L$\uparrow$ vanishes even though 
the AC voltage is applied there.
Figure \ref{fig2} shows the rectified current Eq.\,(\ref{Ianalytic}) as a function of the drive frequency and amplitude.

\begin{figure}[t]
\centering
\includegraphics[scale=0.61]{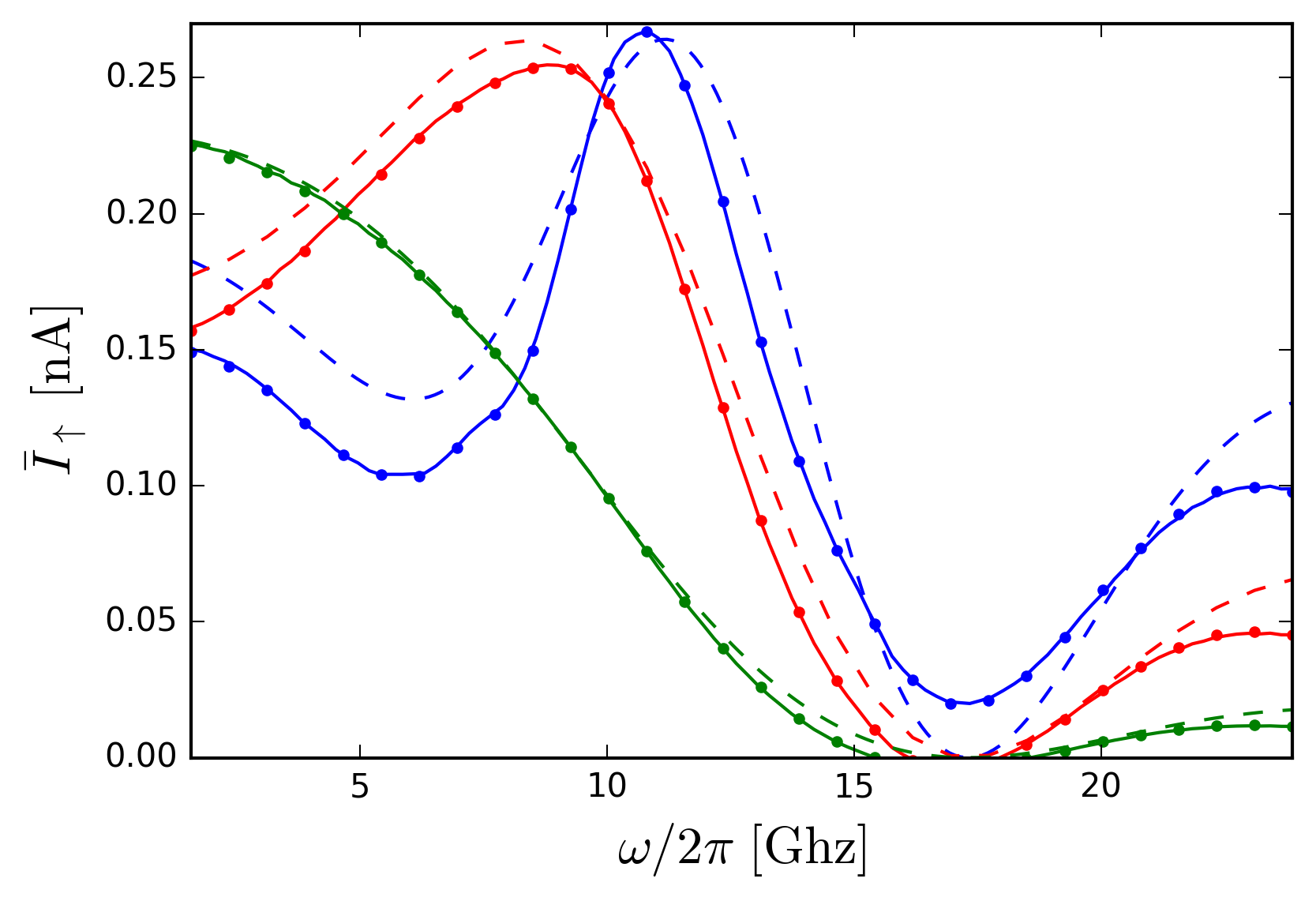}
\vspace{-6ex}
\caption{(Color online) Simple microscopic model. 
 DC current $\bar I_\uparrow$ for three different voltage amplitudes $V(t)=V_0\cos(\omega t)$ with $V_0=31,\ 62,\ 93\,\mu$V (green, red, blue). The symbols correspond to time-dependent simulation of Eq.\,(\ref{eq:H}), the straight lines to semi-analytic theory Eq.\,(\ref{semi-analytical}) and the dash lines to the analytic approach Eq.\ (\ref{Ianalytic}).
}
\label{fig4}
\end{figure}

The DC current follows damped oscillations with both $V_0$ and $\omega$ with frequency $h/\tau$ in the $\sim 10$\,GHz range. 
In particular, the characteristic time $\tau$ can be extracted directly from the minima of the DC current as a function of $\omega$. 
Fig.\,\ref{fig4} shows the plot of current $\bar I_\uparrow$ versus frequency $\omega$ for three different values of $V_0$, corresponding to cuts in Fig.\,\ref{fig2} (green lines). 
Fig.\,\ref{fig4} contains the results of three different calculation: the ideal analytical calculation Eq.\,(\ref{Ianalytic}), the time dependent simulations of the microscopic model Eq.\,(\ref{eq:H}) and a semi-analytical calculation that uses the time independent part of the microscopic model and compute the rectification properties using Eq.\,(\ref{semi-analytical}). 
We find that a close agreement between the three approaches with a very accurate agreement between the latter two. Departure from the ideal analytic formula (\ref{Ianalytic}) arises due to the presence of a small backscattering in the device (which is not perfectly adiabatic) and the fact that the linear relation Eq.\,(\ref{linearphi}) is not strictly valid in the microscopic model (presence of the other characteristic scales).

We conclude that the ideal analytical model Eq.\,(\ref{Ianalytic}) describes the physics qualitatively but cannot be used for quantitative predictions. On the other hand, Eq.\,(\ref{semi-analytical}) is computationally affordable and in precise agreement
with the direct integration of the Schr\"odinger equation. It may be used for other - more realistic - models, which we shall do in the next section.

\section{Realistic microscopic model}

The two models studied above are of course idealized. Below, we develop a much more refined model which builds upon our previous work \cite{Bautze2014}. The model of Ref. \cite{Bautze2014} was shown to be in remarkable agreement with the DC experimental data even though the electrostatic potential was modeled rather crudely. Here, we extend the modelisation and perform a self-consistent treatment of the electrostatic-quantum problem. We also include finite temperature thermal smearing ($\sim 20$\,mK). 

 \begin{figure}[t]
\centering
\includegraphics[scale=0.45]{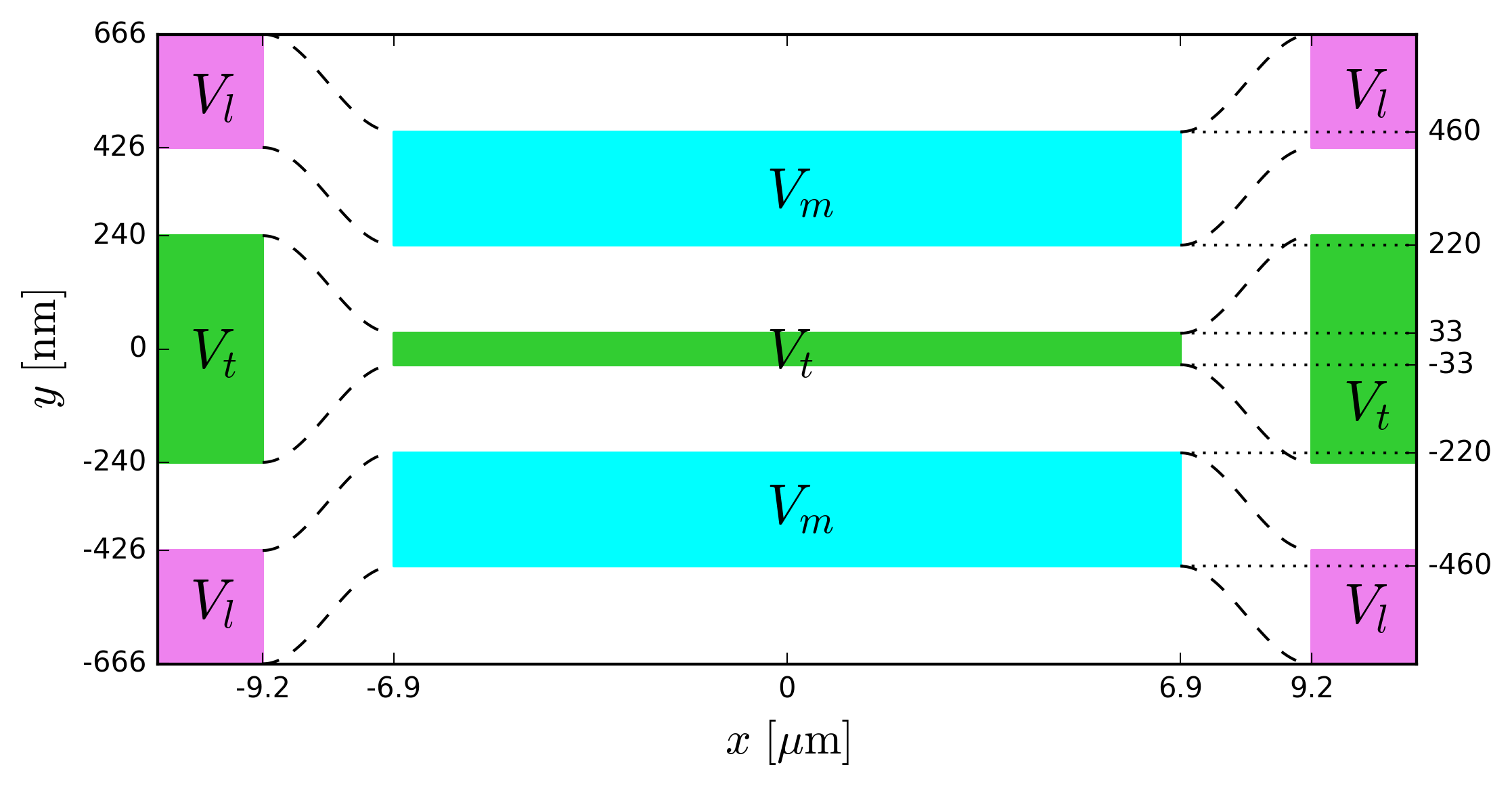}
\vspace{-3ex}
\caption{(Color online) Top view of the layout of the gates that define the ``realistic microscopic model''.}
\label{fig5}
\end{figure}

Before describing the specifics of the "realistic model", let us brifly discuss some orders of magnitude. The typical value of the difference of the time of flight $\tau$ that can be reached experimentally depends on the product of three factors, $\tau=\mathcal{L} \left(1/v_S-1/v_A\right) \approx \mathcal{L}/v_S\times (k_S-k_A)/k_S$. The longitudinal velocity $v_S$ can be estimated from the experimental results of Ref.\,\cite{Roussely2018} to be
$v_S\approx$ 2 - 5\,(10$^5$m\,s$^{-1}$). Typical values of $k_S-k_A$ found in Ref.\,\cite{Bautze2014} lie between $1\%$ and up to $10\%$ of the Fermi momentum $k_S$. The length $\mathcal{L}$ of the tunneling region in Ref.\,\cite{Bautze2014} was $\mathcal{L}= 1\,\mu$m but coherent oscillations have since been observed in much longer samples \cite{Bauerle_priv} $\mathcal{L}\approx 40\,\mu$m indicating that the low temperature ($\approx 20$\,mK) phase coherence length in these samples is of a few tens of $\mu$m, comparable to what has been observed in the quantum Hall regime \cite{Roulleau2008}. Altogether, we  estimates $\tau \sim 100$\,ps for the slowest mode of a $20\,\mu$m long sample, which is consistent with what is found below in the simulations of the realistic model.

\subsection{Geometry}

\begin{figure}[t]
\centering
\includegraphics[scale=0.38]{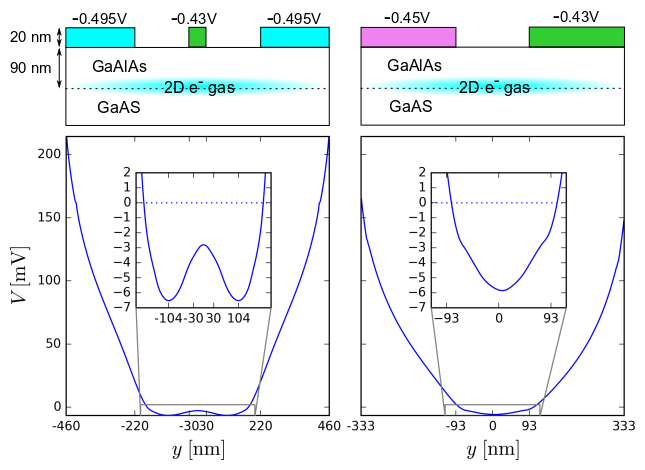}
\vspace{-3ex}
\caption{(Color online) Upper panels: side view of the ``realistic microscopic model'' layout. Lower panels: self-consistent electrostatic potential seen by the electrons as a function of the transverse direction $y$. The insets show a zoom close to the Fermi level $E_F=0$.
Left panels: cut inside the central region ($x=0$). Right panels: cut inside the leads ($x>10\,\mu$m or $x<10\,\mu$m)}
\label{fig6}
\end{figure}

The model is defined solely by the position of the top gates that are deposited on the surface of the GaAs heterostructure. It consists of a central region (defined by two lateral gates and a central tunneling gate) which smoothly evolves into two disconnected wires on the left and on the right of the central region. A top view of the layout of the gates is shown in Fig.\ \ref{fig5}. A cut at $x=0$ (left panel) and $x > 10\,\mu$m (right panel) is shown in the upper panel of Fig.\ref{fig6}.

The dimensions of the device (with a central region 13.8$\,\mu$m long and 0.92$\,\mu$m large) are fully compatible with standard e-beam lithography techniques. The different gates are grouped into three categories: the three interior gates (green) are set to the same potential $V_t$,  the two outer gates of the central region are set to $V_m$ and the four outer gates of the electrodes are set to $V_l$. The transition region between the central region and the lead ($x\in [-9.2,-6.9]$ and $x\in [6.9,9.2] \ \mu$m ) is defined by an interpolation described later in this section.

\subsection{Self-consistent model}

In order to calculate the electrostatic potential seen by the two-dimensional electron gas, we work in the effective mass ($m^* = 0.067\,m_e$, $m_e$: bare electron mass) approximation for
the Schr\"odinger equation which is solved self-consistently with Poisson equation. The Hamiltonian of the two dimensional electron gas,
\begin{equation}
H = \frac{P^2_x+P^2_y}{2m^*} - e V(x,y, z=0),
\end{equation}
is discretized on a square grid with lattice constant $a = 3$\,nm (approximately 2$\times10^6 \approx 300\times 6000$ sites).
The Schr\"odinger equation 
\begin{equation}
H \Psi_{\alpha E} = E \Psi_{\alpha E}
\label{schro}
\end{equation}  
is solved using the Kwant package\cite{Groth2014}. The electrodes are taken to be semi-infinite so that the spectrum is actually continuous and the eigenfunctions labeled by an energy $E$ and a mode index $\alpha$. The density of electrons $n(x,y)$ is given by the integral over energy of the local density of states,
\begin{equation}
\label{dst}
n(x,y) = \sum_\alpha \int \frac{dE}{2\pi} |\Psi_{\alpha E}(x,y)|^2 f(E)
\end{equation}
where $f(E)=1/(e^{E/k_BT} +1)$ is the Fermi function at temperature $T$ (and we have set the Fermi energy $E_F=0$ as our reference energy point). The Poisson equation away from the electron gas reads
\begin{equation}
\label{P1}
\Delta V(x,y, z) =0
\end{equation}
while close to the gas the discontinuity of the electric field is set by $n(x,y)$:
\begin{equation}
\label{P2}
\partial_z V(x,y, 0^+) - \partial_z V(x,y, 0^-) = - \frac{e}{\epsilon} [ n(x,y) + n_d]   
\end{equation}
where the dopant density $n_d$ sets the actual density of the gas and $\epsilon\approx 12\epsilon_0$ is the dielectric constant. 
The Poisson equation is solved using the FEniCS package \cite{LoggMardalEtAl2012a}. 

In order to solve  the set of self-consistent equations
(\ref{schro}, \ref{dst}, \ref{P1}, \ref{P2}), we perform one approximation which considerably lowers the 
computational effort while retaining good accuracy. 
In a first step, we solve the self-consistent problem deep in the lead region where the system is
invariant by translation along $x$ (hence effectively maps onto a 2D problem for the Poisson equation and 1D for the quantum problem). We obtain $V(|x|\gg 10, y, 0)\equiv V_A(y)$.  Secondly, we solve the problem deep inside the central region, assuming that the potential is not affected by the leads (hence also invariant by translation along $x$). We obtain $V(|x|\ll 10, y, 0)\equiv V_B(y)$. An example of the obtained self-consistent potentials $V_B(y)$ (left) and $V_A(y)$ (right) is shown in
Fig.\ \ref{fig6} for $V_t = -0.43$\,V, $V_m=-0.495$\,V and $V_l = -0.45$\,V. In the last step, we describe the potential in the
transition regions ($x\in [-9.2, -6.9]$ and $x\in [6.9, 9.2]$ ) by performing an interpolation between $V_A(y)$ and 
$V_B(y)$. 

\begin{figure}[t]
\centering
\includegraphics[scale=0.48]{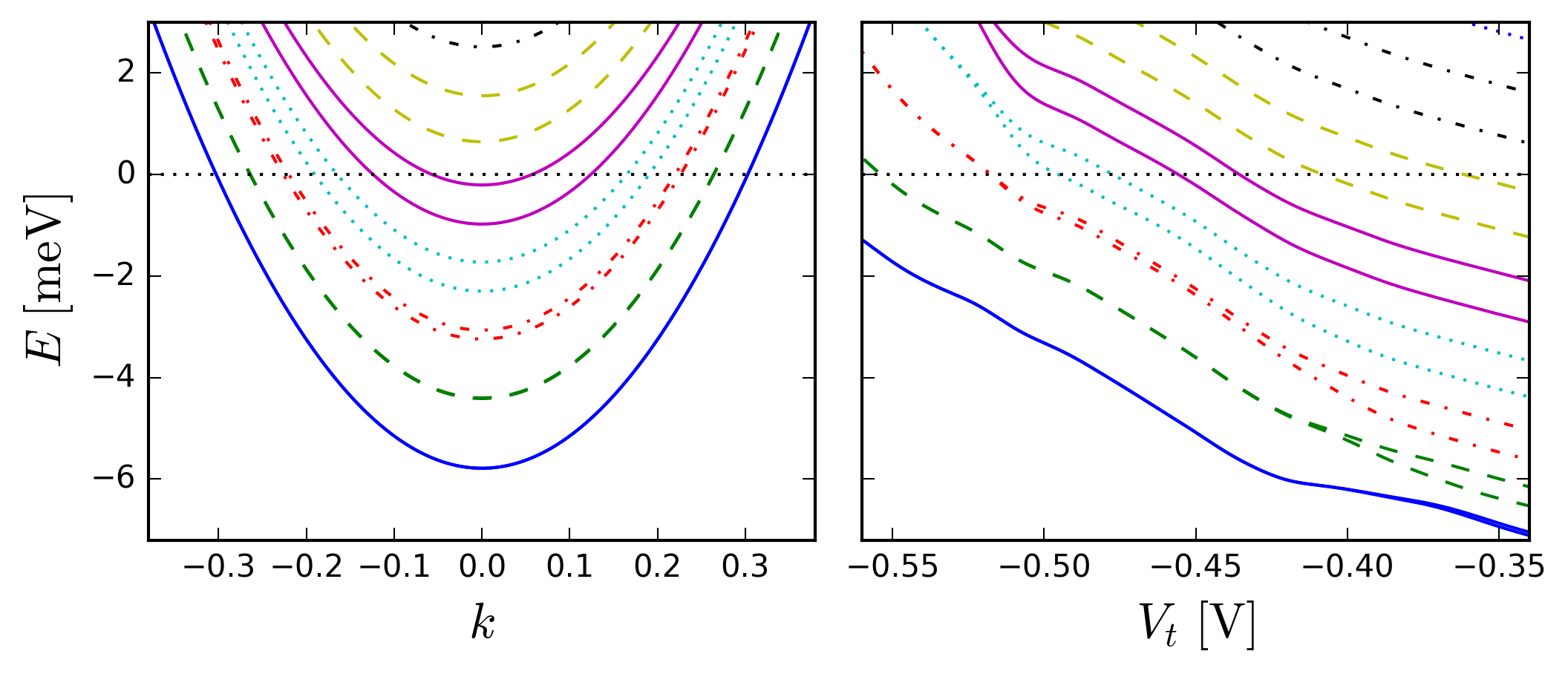}
\vspace{-3ex}
\caption{(Color online) Structure of the subbands in the central region. Left panel: Energy dispersion $E(k)$ versus $k$ for $V_t=-0.43$\,V. The bands that cross the Fermi energy $E=E_F=0$ correspond to propagating channels. Right panel: Transverse energies $E(k=0)$ of the different modes as a function of the tunneling voltage $V_t$. The bands below the Fermi energy are propagating. Parameters: $V_m=-0.495$\,V and $V_l=-0.45$\,V for both panels. Symmetric/Antisymmetric modes pairs are plotted with similar color and line style.}
\label{fig7}
\end{figure}

The density of the gas is $\sim 3.2\times10^{11}$\,cm$^{-2}$ which corresponds to a Fermi wave length $\lambda_F \approx 45$\,nm. Since the transition region is long compared to $\lambda_F$, the transition is adiabatic and we observe very small reflexion probability. Who also have to check that mode coming from $L_\uparrow$ are fully transmitted into mode of $R_\uparrow$ and $R_\downarrow$ and no leak into $L_\downarrow$. Otherwise it would create Fabry-Perot interferences between both potential transitions which will compete with the Mach-Zehnder interferometry. This is achieve by a smooth transition of the potential.

\subsection{DC and AC characterization}

\begin{figure}[t]
\centering
\includegraphics[scale=0.48]{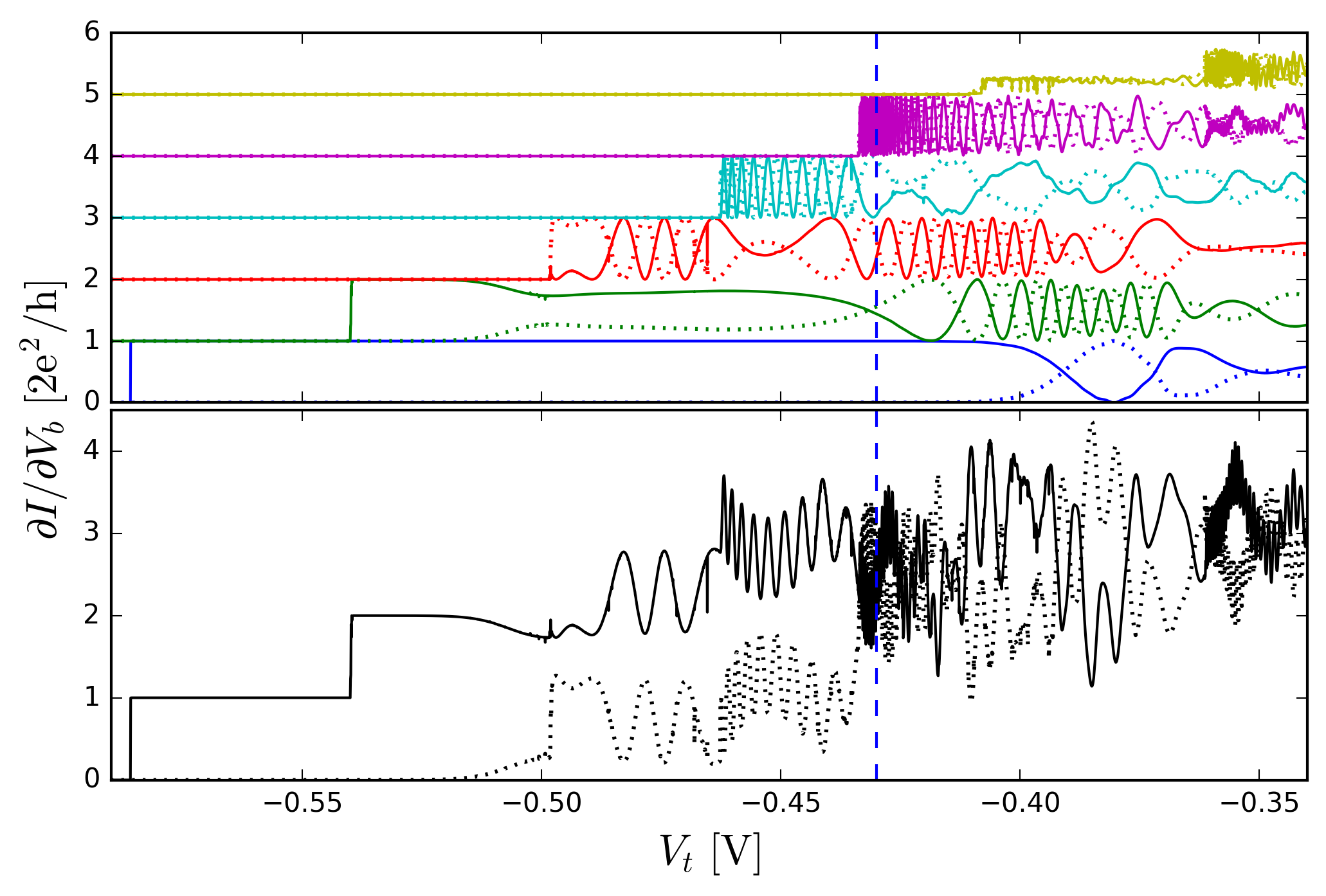}
\vspace{-3ex}
\caption{(Color online) Lower panel: DC differential conductance $\partial I/ \partial V_b$  as a function of the central gate voltage $V_t$ for $V_l=-0.45\,$V and $V_m=-0.495\,$V. The voltage bias $V_b$ is applied to the upper left contact $L_\uparrow$ while the other three are grounded. The current is measured in the upper right contact $R_\uparrow$ (full lines, $\partial I_{\uparrow}/ \partial V_b$) and in the lower right contact $R_\downarrow$ 
(dashed line, $\partial I_{\downarrow}/ \partial V_b$). Upper panel: contribution from the individual propagating channels, shifted by multiples of $2e^2/h$ for clarity. Calculations performed at zero temperature}
\label{fig8}
\end{figure}

Once the electrostatic potential is known, we calculate the transmission probabilities for the various conducting channels. We have used $V_t=-0.43\,$V, $V_l=-0.45\,$V and $V_m=-0.495\,$V so that five propagating channels are open in each lead, and ten channels are open in the central region (a typical experimental situation). The left panel of Fig.\ \ref{fig7} shows an example of band structure of the central region where we have used matching colors to identify the symmetric/antisymmetric pairs. The right panel of Fig.\ \ref{fig7} shows $E(k=0)$ for the various modes which allows one to identify the propagating channels ($E(k=0)<0$) and evaluate the splitting between the symmetric and antisymmetric components. 
Fig.\ \ref{fig8} shows the DC conductance (at zero temperature) as a function of the tunneling voltage $V_t$ (lower panel) obtained with Kwant\cite{Groth2014}. The upper panel shows the contributions from the different propagating channels. The strongest oscillating signal is obtained close to the onset of the opening of a new channel where the two momenta for the symmetric and anti-symmetric channels are the most different.

\subsection{Rectification spectroscopy}
\begin{figure}[t]
\centering
\includegraphics[scale=0.47]{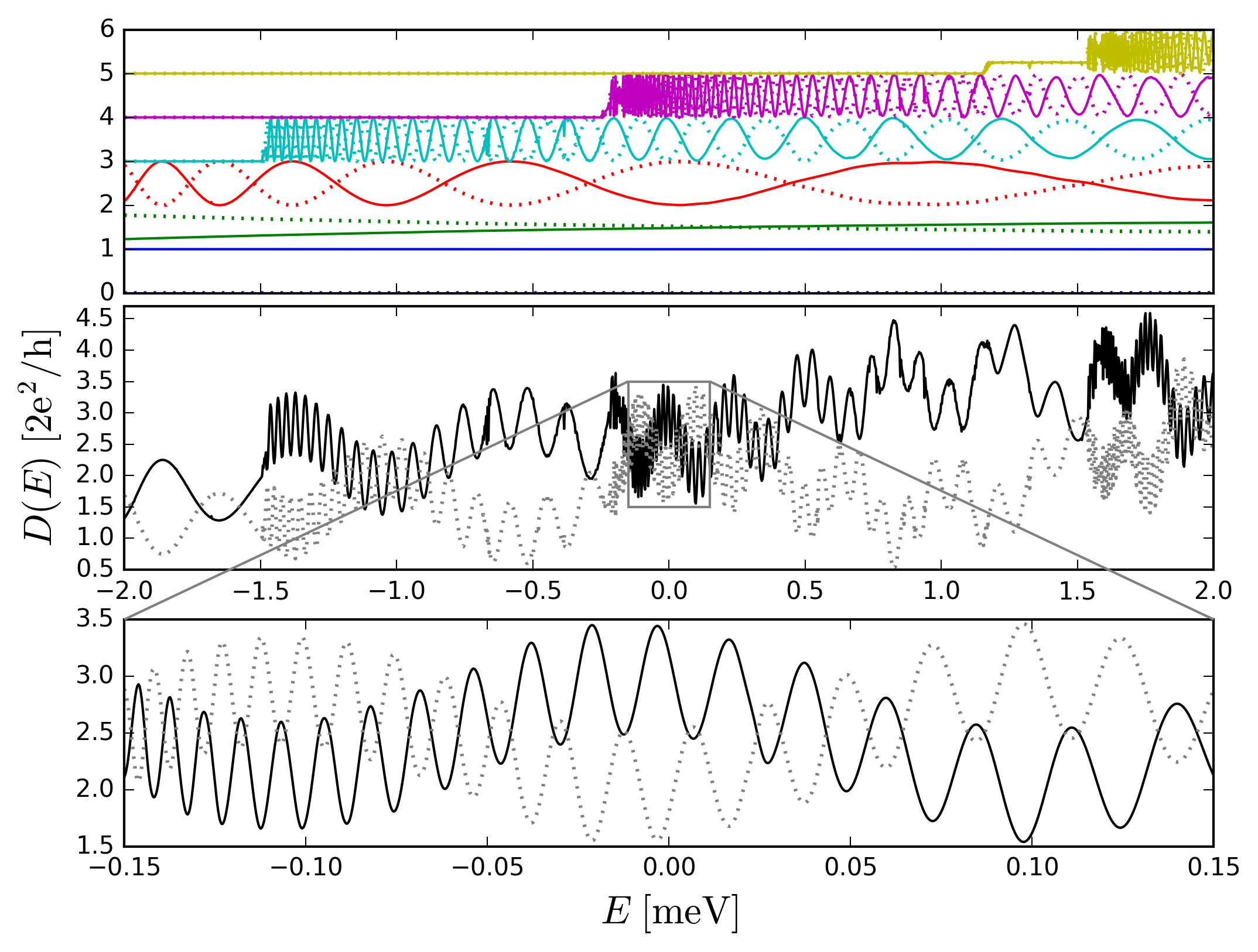}
\vspace{-3ex}
\caption{(Color online) Middle panel: Total transmission probability $D(E)$, [full lines, $D_{\uparrow \uparrow}(E)$, dashed line, $D_{\uparrow \downarrow}(E)$] vs.\ energy $E$ , where $E$ is measured relatively to the Fermi energy $E_F$. Upper panel: contribution from the individual propagating channels, shifted by multiples of $2e^2/h$ for clarity. Lower Panel: zoom of the middle panel.
Parameters: $V_t=-0.43$\,V, $V_l=-0.45\,$V and $V_m=-0.495\,$V for all panels.}
\label{fig9}
\end{figure}

The total number of orbitals is now rather large ($\sim 2\times10^6$) so that a direct time dependent calculation is prohibitive. But discussion of section \ref{comparison approach} shows that we can use Eq.\ (\ref{semi-analytical}) and get the same results with much less computational time. In order to obtain the rectification current one requires the calculation of the total transmission probability 
\begin{equation}
D_{ab}(E) = \sum_{\alpha\in a,\beta\in b} |d_{\alpha\beta}(E)|^2
\end{equation}  
where  the sum is take onto all the propagating channel of the corresponding electrode. An example of such a calculation using Kwant\cite{Groth2014} is shown in Fig.\ \ref{fig9} together with the detailed contributions of the different channels. The curve $D_{ab}(E)$, antisymmetrized around the Fermi level, provides the information for the calculation of the rectified current response to an AC drive as can be seen from the following reformulation of  Eq. (\ref{semi-analytical}),
\begin{align}
\label{semi-analytical-AS} 
\bar I_b = \frac{e}{h} \sum_{n>0} |P_n|^2 \int dE 
&\left[D_{ba}(E_{-n}) - D_{ba}(E_{n}) \right] \nonumber\\
&\times\left[ f_a(E_{n})-f_a(E_{-n})\right]
\end{align}
where $E_n = E + n\hbar\omega/2$. 
Conversely, Eq.(\ref{semi-analytical-AS}) shows that the rectification response can be used to reconstruct the anti-symmetrized transmission probability of the device. To reconstruct the full transmission probability, including the symmetric part, calculations/measurements for different Fermi levels (using e.g. a back gate) are necessary.

\begin{figure}[ht]
\centering
\includegraphics[scale=0.61]{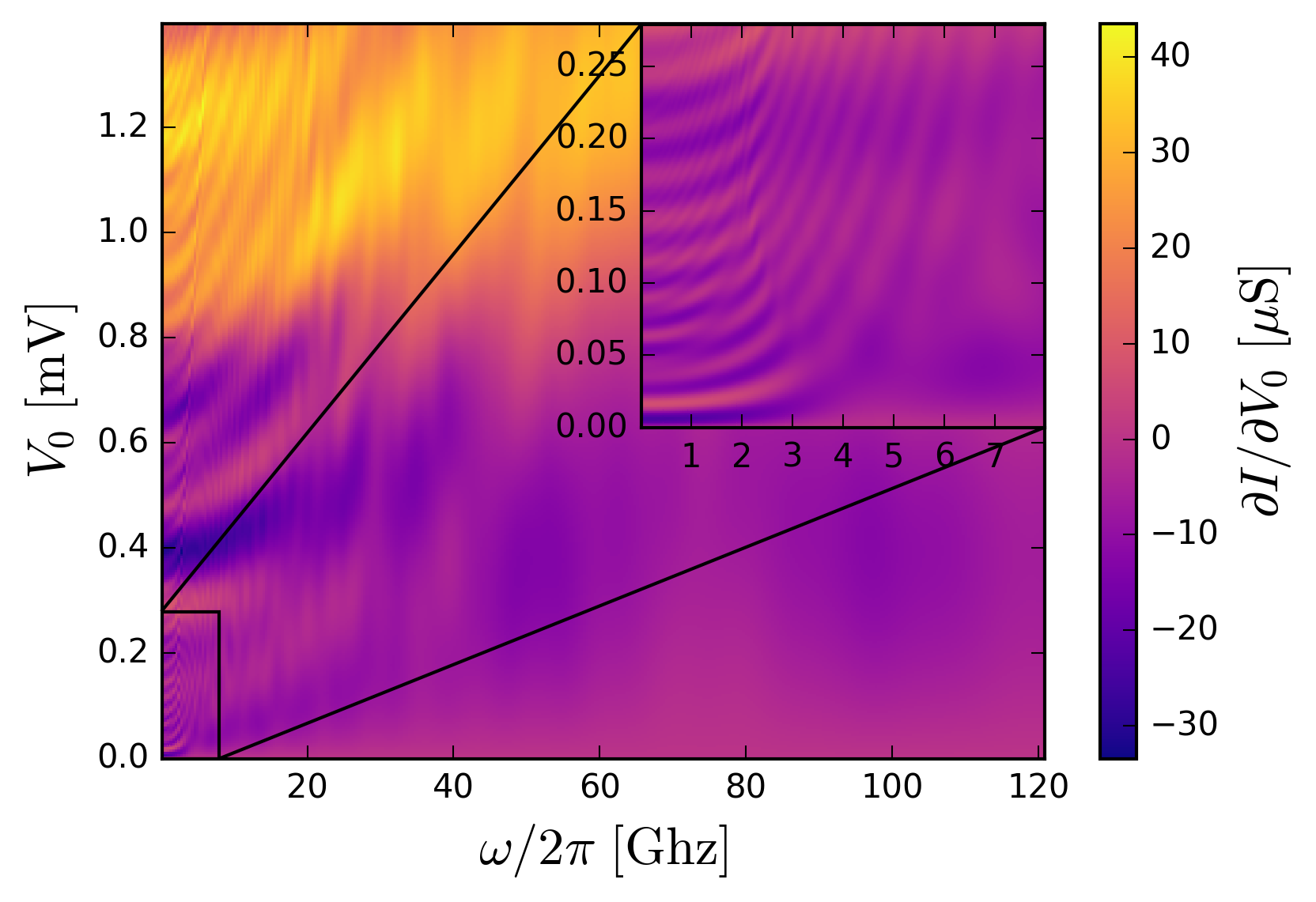}
\vspace{-6ex}
\caption{(Color online) Realistic microscopic model. Colormap of $\partial\bar{I_\uparrow}/\partial V_0 -\partial\bar{I_\downarrow}/\partial V_0$ versus voltage amplitude $V_0$ and frequency $\omega/2\pi$. The inset shows a zoom of the main panel. Two channels with $\tau = 220$ps (oscillations visible in the inset) and $\tau = 19$ps (oscillations visible in the main panel) dominate the signal. Calculations performed at 20 mK.}
\label{fig10}
\end{figure}

The resulting rectified current for the realistic model is Fig.\ \ref{fig10}. Fig.\ \ref{fig10} is qualitatively similar to the idealized model despite the fact that it includes a realistic modeling of the electrostatic potential, multiple opened channel ($5$) and a finite temperature (20 mK). This is a strong indication of the robustness of this type of spectroscopy.

An important aspect of the multi-channel model is that different channels (with different scales $\tau$)
contribute to the rectified current with contributions of order $1/\tau$ so that the fastest channels
have larger contributions. However, this does not prevent one from observing the slowest channels since the
scales at which the different contributions vary is also very different [as can be inferred by an inspection of Eq.(\ref{Ianalytic})]. In order to bring the different contributions to the same scale, it can be advantageous to
plot the derivative of the current $\partial \bar I/\partial V_0$ instead of the current itself. This is typically performed experimentally using a lock-in technique. The signal can be furthered amplified by plotting the anti-symmetric signal $\partial\bar{I_\uparrow}/\partial V_0 -\partial\bar{I_\downarrow}/\partial V_0$ with respect to the two outputs in order to subtract any spurious signal coming from other rectification processes. Indeed, the multi-channel realistic model contains another source of rectification current coming from the opening of new channels which give rise to plateaus in the rectification current. These plateaus are very conveniently subtracted by looking at the anti-symmetric signal $\partial\bar{I_\uparrow}/\partial V_0 -\partial\bar{I_\downarrow}/\partial V_0$.

The data of Fig. \ref{fig10} corresponds to 5 pairs of propagating channels with respectively $\tau = 220$\,ps, $19$\,ps, $\tau = 3$\,ps and two very fast channels with $\tau\ll 1$\,ps. With curent experimental capabilities, the two interesting pairs that may be used for flying qubits are  the two slowest $\tau = 220$\,ps and $19$\,ps. It is interesting that despite the presence of the three faster pairs, the spectroscopy lines of these two pairs are clearly visible in Fig. \ref{fig10}: at these scales, the three fast pairs only contribute to a global background. The two characteristic times $\tau = 220$\,ps and $19$\,ps can be directly extracted by fitting the low frequency ($< 10$ GHz) and large frequency ($< 100$GHz) part of the diagram.

\section{Discussion and Conclusion}

The experimental observation of the features shown in Fig.\ \ref{fig10} would provide the first direct measure of the characteristic times of the device and validate the possibility for the dynamical probing of an interference pattern at high frequency. This is a key step on the route toward further quantum manipulation with voltage pulses and the first full fledged electronic flying qubit\cite{BAUERLE2018}.

Another important aspect which is at stakes is our ability to make accurate models, and predictive simulations, for high frequency quantum transport. At the experimental level, the electrostatic gates are controlled with voltages of the order of 1 V while the equilibrium electrostatic potential seen by the electrons is of the order of several mV, i.e. 2-3 orders of magnitude smaller (see e.g. Fig.\ref{fig6}). Hence, the construction of accurate models  must go through a precise understanding of the combined electrostatic-quantum problem in presence of high frequency dynamics. Conversely, the physics of these systems depends on the precise interplay between these two physics. Being in position to make quantitative predictions for these systems would allow one to design much more optimum geometries and experimental protocols; it would have a decisive impact in the development of the field. This article presents a step in this direction. 

Our understanding of high frequency quantum transport, pulse propagation and dynamical interferometry (the ingredients of electronic flying qubit architectures) is mostly based so far on non-interacting models. As the experiments progress toward the exploration of this new physics, the modeling will require new aspects to be treated more accurately. Future work shall include
a proper treatment of the electron-electron interactions at the RPA level\cite{Kloss2018}.   and beyond as well
as a the modelisation of the different channels for decoherence. Indeed, understanding what sets the fundamental limit of coherence in these systems will probably be one of the most interesting challenge of the field in the years to come.

{\it Acknowledgment.} This work was supported by the ANR Full Quantum, the ANR QTERA, the French-Japon ANR QCONTROL and the US Office of Naval Research. We thank Chris B\"auerle for useful discussions.

\bibliographystyle{apsrev4-1_own}
\bibliography{biblio}

\end{document}